# Revealing the Pulmonary Surfactant Corona on Silica Nanoparticles by Cryo-Transmission Electron Microscopy

Fanny Mousseau,[*,a] Evdokia K. Oikonomou,[a] Annie Vacher,[b] Marc Airiau,[b] Stéphane Mornet[c] and Jean-François Berret[a]

[a)]*Laboratoire Matière et Systèmes Complexes, 10 rue Alice Domon et Léonie Duquet, 75205 Paris Cedex, France. E-mail: fanny.mousseau@univ-paris-diderot.fr*
[b)]*Solvay Research & Innovation Center Paris, 52 rue de la Haie Coq, 93306 Aubervilliers Cedex, France*
[c)]*Institut de Chimie de la Matière Condensée de Bordeaux, UPR CNRS 9048, Université Bordeaux 1, 87 avenue du Docteur A. Schweitzer, Pessac cedex F-33608, France*

**ABSTRACT. When inhaled nanoparticles (NPs) deposit in alveoli, they transit through the pulmonary surfactant (PS), a biofluid made of proteins and phospholipidic vesicles. They acquire a corona reflecting the PS-nanomaterials interaction. Since the corona determines directly the NPs biological fate, the question of its nature and structure is central. Here, we report on the corona architecture raising after incubation of positive or negative silica particles with Curosurf®, a biomimetic pulmonary surfactant of porcine origin. Using optical, electron and cryo-electron microscopy (cryo-TEM), we determine the pulmonary surfactant corona structure at different scales of observation. Contrary to common belief, the PS corona is not only constituted by phospholipidic bilayers surrounding NPs but by multiple hybrid structures derived from NPs-vesicles interaction. Statistical analysis of cryo-TEM images provides interesting highlights about the nature of the corona depending on the particle charge. The influence of Curosurf® pre- or post-treatment is also investigated and demonstrate the need of protocols standardization.**

Nanoparticles (NPs) in contact with biological fluids are rapidly covered by selected groups of biomolecules including proteins, lipids, specialized sugars, ions and enzymes, to form a corona that modifies nanomaterials interaction with cells and tissues.[1,2] Following the work performed on protein corona in blood, research has started to address the problem of the corona formation in other biofluids, as for instance pulmonary surfactant (PS). When inhaled, nanoparticles may reach the alveolar space and come in contact with a thin layer (< 1 µm) of the fluid lining the alveolar epithelium: the pulmonary surfactant.[3] This fluid is made of phospholipids (90%) and proteins (10 %) and its total concentration is estimated at 40 g L$^{-1}$.[3] At the air-liquid interface, phospholipids are assembled into mono- and multi-layers while beneath the surface, they form multilamellar vesicles and tubular myelin.[3] Upon contact with PS, a corona of proteins and / or lipids is formed at the particle surface.[4] Such a pulmonary surfactant corona (PS corona) is different from the one found with serum proteins in plasma.

The physicochemical characteristics of NPs, i.e. their chemical composition, size and surface charge, have been shown to notably affect NPs interactions with lung fluids and to modify their cellular pathway, toxicity and uptake.[5–7] Studying the PS corona for different nanomaterials is a major challenge in the fields of nanotoxicology and nanomedicine. Despite their importance for lung physiology, studies dealing with the PS corona are still scarce. They mainly rely on mass spectrometry results and on the determination of the corona chemical composition but provide few information about its structural organization.[4,8–12] Recent works focus on the modification of NPs – cells interactions in presence of pulmonary surfactant.[5,6,13–19] However, both mechanisms by which the PS corona is formed and its biological consequences remain open questions. Solving these issues require *in vitro* and *in vivo* experiments under various conditions.

Pertaining to nanomaterials bioreactivity in alveolar spaces, studies have shown that it is important to characterize their corona at the point of exposure. To this end, pioneering works investigate the impact of hydroxyapatite and titanium oxide nanoparticles on PS function and structure using transmission electron microscopy (TEM). NPs are shown to aggregate in solution or to be incorporated between PS bilayers, leading to a vesicular size reduction.[20,21] However, the corona architecture could not be determined precisely because artefacts due to conventional staining or drying processes are generally present in TEM studies of assembled structures such as micelles or vesicles.[22]

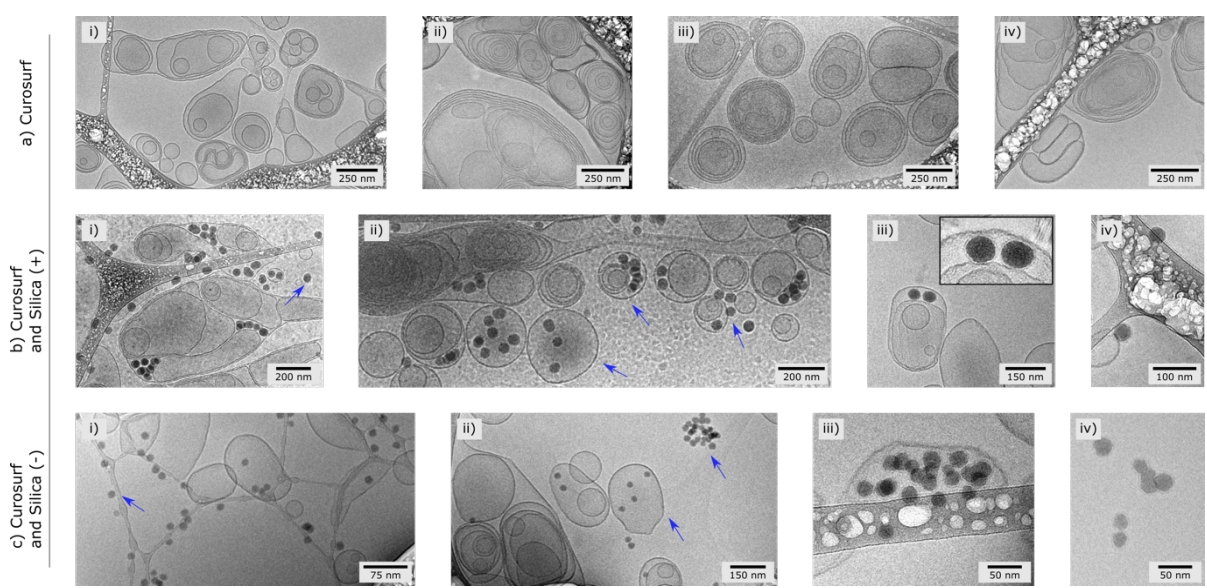

**Figure 1.** *Cryo-TEM pictures of Curosurf® (a), Curosurf® and Silica (+) NPs (b) and Curosurf® and Silica (-) NPs (c). Additional images are provided in **S4**.*

The 2017 Nobel Prize in chemistry was awarded to Jacques Dubochet, Richard Henderson, and Joachim Frank, for "developing cryo-electron microscopy for the high-resolution structure determination of biomolecules in solution". *De* facto, cryo-TEM, the first reference method for the structural analysis of specific biological samples,[22] was used for imaging the PS corona structure. Kumar *et al.* show that silica or polymeric nanoparticles incubated in PS lead to the spontaneous formation of a supported lipid bilayer (SLB)[10] whereas Theodorou *et al.* demonstrate the formation of multilayers of phospholipids on silver and zinc oxide nanowires after putting them in contact with PS.[18,23] Finally, Konduru *et al*. observe that cerium oxide and barium sulfate NPs incubated with PS form hybrid NPs-vesicles agglomerates.[24] Thus the few available studies, which by the way display only zoomed images of some nanoparticles, lead to contradictory conclusions about the corona architecture. Concerning simulation, recent reports examine the spatial organization of lipid membranes with respect to the NPs. For instance, Hu *et al.* consider PS vesicles of complex lipid and protein compositions and show that their interaction with silver or polystyrene NPs led to mono- or bilayer coating formation where proteins are also present.[25]

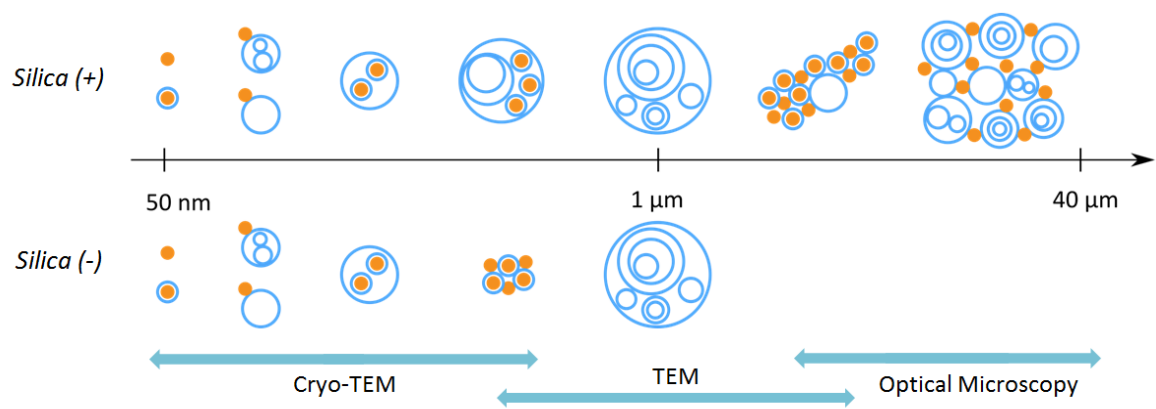

**Scheme 1.** *Representation of the hybrid colloids present in the PS corona of positive and negative silica nanoparticles. The scales covered by the different techniques of microscopy used in this work are displayed in blue.*

Here, we report the first statistical analysis of the pulmonary surfactant corona structure associated with nanoparticles. The NP / lipid assembly is investigated via optical microscopy, TEM and cryo-TEM for positive and negative silica particles, allowing to draw the global picture of the PS corona. In addition, we study the impact of the formulation pathway on the corona structure (materials and methods are detailed in Supplementary Information S1).

Curosurf® is the pulmonary surfactant considered in this work. This PS is extracted from minced pork lungs and is used for the surfactant replacement therapy of premature infants. Its lipids and proteins compositions are provided in S1 and compared to those of human surfactant. Figure 1a displays the Curosurf® ultra-structure. It is composed of uni- and multi-lamellar vesicles disperse in size, similarly to the human PS. In our previous work, we provided an extensive characterization of Curosurf® physico-chemical properties.[26] In particular, we showed that Curosurf® vesicles exhibit a bimodal size distribution centered on 80 and 800 nm, have a zeta potential of – 55 mV in MilliQ water and are made of 4.3 nm thick phopholipic bilayers present in the fluid state at 37 °C (S2). Amorphous silica particles have been selected due to their potential cytotoxicity when inhaled as dust.[27,28] Indeed, silica inhalation can cause lung-related diseases including inflammation, silicosis and lung cancer. Positive and negative NPs of respectively 40 and 20 nm are used here to investigate the role of the nanoparticle charge on the PS corona structure. In the following, they are abbreviated Silica (+) and Silica (-). In S1, the physico-chemical properties of these particles are summarized. In particular their charge densities are determined at +0.62 and -0.31 e.nm$^{-2}$ respectively.

| Table 1.1 | NPs on vesicles | NPs inside vesicles | NPs in aggregates | Isolated NPs |
|---|---|---|---|---|
| Curosurf / Silica (+) | 10 % | 63 % | 10 % | 17 % |
| Curosurf / Silica (-) | 4 % | 43 % | 29 % | 24 % |
| Extruded Curosurf / Silica (+) | 3 % | 10 % | 84 % | 3 % |
| Curosurf / Silica (+) after sonication | 0 % | 0 % | 0 % | 100 % |

| Table 1.2 | Liposome | Multilamellar vesicle |
|---|---|---|
| Curosurf / Silica (+) | 45 % | 55 % |
| Curosurf / Silica (-) | 68 % | 32 % |
| Extruded Curosurf / Silica (+) | 85 % | 15 % |

**Table 1.** *Relative amount of sub-micron hybrid colloids present in the different PS coronas.*

In a previous work, we demonstrated via Job scattering plots[29] that the interaction of silica nanoparticles and Curosurf® is mainly driven by electrostatics.[30] Here, we aim at disclosing the PS corona structure at the nano- and micrometer scales using a set of different techniques. First, optical microscopy is performed on Silica (+) or Silica (-) incubated with fluorescent Curosurf®[31] (S3.1 and S3.2). Images show that pulmonary surfactant and Silica (+) interaction gives rise to large micrometer sized hybrid aggregates. Meanwhile, no such aggregates are found with Silica (-). TEM performed on Silica (+) / Curosurf® systems reveals that the aforementioned aggregates are of two kinds (Scheme 1 and S3.3). On the one hand, the particles act like stickers that bind the vesicles together via a physical link. On the other hand, hybrid aggregates contain pristine vesicles, bare particles and particles covered by a SLB. Our results suggest that the PS corona cannot be only described in terms of supported lipid bilayers and that a refined picture is necessary.

Cryo-TEM images in Figure 1b-c and S4 show silica particles incubated with Curosurf®. Mixed silica–surfactant dispersions were prepared at concentrations inducing a surface ratio $X_S$ = 75, where $X_S$ denotes the ratio between the specific surface associated with the vesicles and that associated with the particles. Therefore, as in physiological conditions, PS is present in large excess and is able to cover all the particles by a SLB. More than 800 NPs are analyzed for each particle type and classified depending on their assembly with phospholipids. Four different hybrid structures are found: NPs are either adsorbed at vesicles outer membrane, internalized inside vesicles, aggregated with other particles or isolated (Scheme 1, arrows in Figure 1 and Table 1.1). As expected, significantly more particles are found internalized in vesicles than adsorbed at their outer surface whatever the NP charge. Indeed, particle internalization occurs if the adhesion energy between the particles and the membrane is greater than the bending energy associated to the wrapping process, e.g above an estimated critical diameter of 20 nm.[32] Given the dispersity of the two silica studied (Table S1.1), they are larger than this critical size.

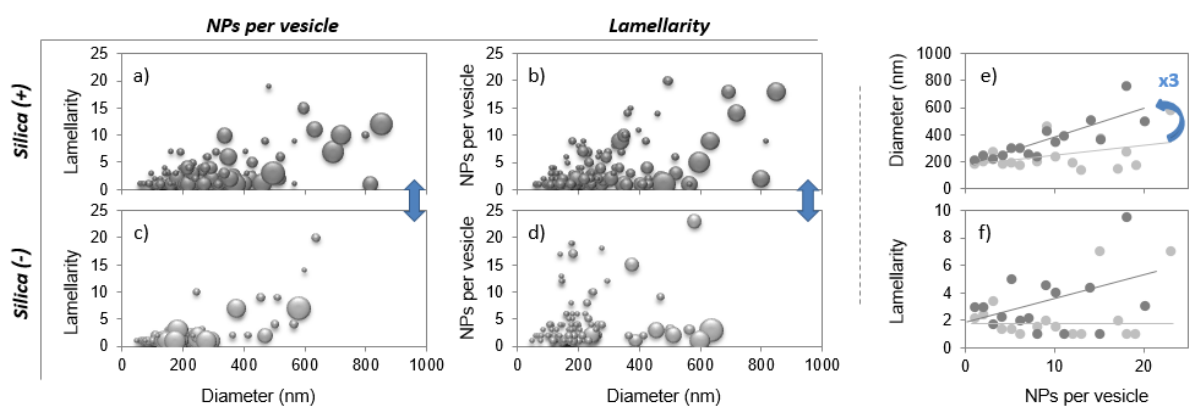

*Figure 2.* *a-d) 3D plots representation of the number of internalized NPs as a function of the vesicles lamellarity and diameter. The bubbles size is proportional to the number of wrapped NPs (a, c) or to the lamellarity of the vesicles (b, d). e-f) Mean diameter and mean lamellarity of vesicles as a function of the number of internalized NPs for positive (dark circles) and negative (bright circles) NPs. In f), lines are guides for eyes.*

Moreover, statistics show that after incubation with Curosurf®, 63 % of the positive NPs are internalized by vesicles whereas this number decreases to 43 % for the negative particles (Table 1.1). In addition, isolated particles are three times more numerous with Silica (-) than with Silica (+). Close up views of these particles show that they are either bare or coated by a SLB (Figure 1b-iv and Figure 1-c-iv). All these results are in agreement with our previous findings which indicated that Silica (+) interact more strongly with Curosurf® than Silica (-) (S3). Table 1 shows that both positive and negative silica particles give similar hybrid structures in contact with PS, but that the relative proportions of the different structures depend on the NP charge (note that a size effect cannot be completely excluded in the results obtained).

Be that as it may, coupling cryo-TEM, TEM and optical microscopy enables us to retrieve a complete description of the PS corona. (Scheme 1). NP – lipid interaction includes van der Waals and electrostatic interactions, hydration forces and bilayers thermal fluctuations[33]. The mechanisms describing these interactions, and consequently the obtained hybrid structures, are thus complex and multifactorial. Hence, the coronas obtained in this work are specific to our experimental conditions, which highlights the general need to carefully characterize the corona formed for each NPs / PS studied system.

Special attention is now drawn on internalized nanoparticles. While Silica (+) are found in uni- or multi-lamellar vesicles in equal numbers (45 % versus 55%), 68 % of Silica (-) are internalized in liposomes (Table 1.2). Regardless of the particle charge and vesicles lamellarity, internalized particles exhibit a SLB[34] (Figure 1b-iii and Figure 1c-iii). In the particular case of NPs inside multilamellar vesicles, particles are always found between the outer and inner membranes. This original engulfing process suggests that only the outer membrane initiates the internalization process and confirms that SLB-coated NPs cannot enter liposomes.[32]

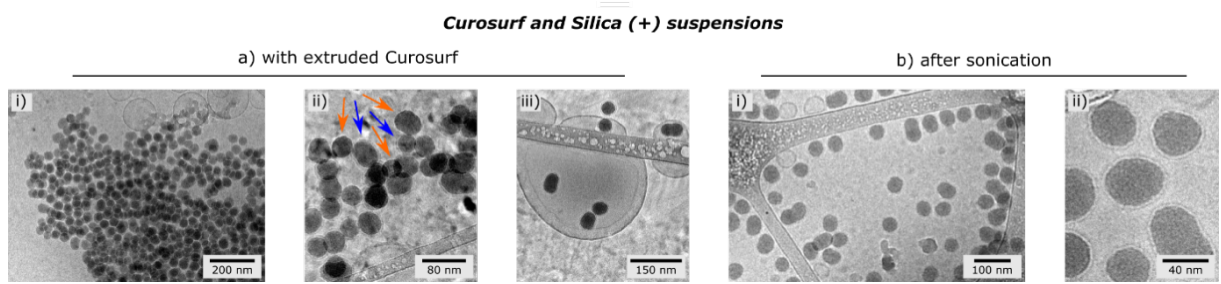

**Figure 3.** Cryo-TEM images of Silica (+) / Curosurf ® systems with extruded Curosurf® (a) or after sonication (b). In a-ii, orange and blue arrows indicate respectively naked and coated NPs

Each uni- and multi-lamellar vesicle containing particles is then analyzed regarding 3 parameters: the number of internalized NPs, the vesicle lamellarity and the vesicle diameter. Results collected on more than 100 vesicles are summarized via 3D plots to establish cross-correlations between the aforementionned parameters. The patterns herein displayed are representative of the PS corona and differ one from another according to the NP charge (Figure 2a-d). Then, the lamellarity and the vesicle diameter are represented versus the number of internalized particles (Figure 2e,f). Results show that up to 23 nanoparticles can be wrapped by vesicles and that on average, this value is not affected by the particle charge. Interestingly, the diameter of vesicles containing NPs varies linearly with the amount of captured NPs, with a slope 3 times higher for the Silica (+). Thus, as vesicles become larger, their lamellarity increases and they are able to invaginate more positive NPs. In contrast, Silica (-) are mainly internalized by liposomes, as indicated in Table 1.2. Our results are in agreement with those of Konduru *et al*. where positive cerium oxide NPs interact mainly with multilamellar vesicles of PS whereas negative barium sulfate NPs interact with liposomes in majority.[24] Definitely, our study shows that the PS corona structure is significantly different when formed on positive or negative silica nanoparticles. This outcome is of major importance since the binding of PS to a variety of nanoparticles has been shown to influence their cellular uptake, intracellular localization, oxidative properties, inflammation potential as well as their toxicity in the lungs.[14,16,18,19]

Finally, the influence of the formulation pathway on the PS corona structure is studied for two reasons. First, liposomes made of synthetic phospholipids and obtained by extrusion are often used as PS substitute in physico-chemical studies of NPs / PS interactions. Second, in *in vitro* cellular assays NPs are usually incubated in PS then sonicated to allow their dispersion before being put in contact with cells.[24,35] In the following, we thus investigate how pre-extrusion of PS and post-sonication of NPs / PS systems affect the corona structure. With this aim, Silica (+) particles are incubated with extruded Curosurf® (vesicle diameter reduced from 800 nm to 130 nm, S5). Cryo-TEM images show that 90 % of the NPs are comprised in large micron sized aggregates (Table 1 and Figure 3a) and that the hybrid assemblies are indeed made of bare and SLB-coated particles (Figure 3a-ii). As these latter are negative[19], aggregation probably results from electrostatic interaction of oppositely charged species. In a second assay, Silica (+) and Curosurf® are incubated then sonicated. 100 % of particles are found non-aggregated and coated with a SLB (Figure 3b) as observed on other systems.[36,37] Our results demonstrate that pre- or post-treatment of Curosurf® such as extrusion and sonication strongly modifies the PS corona.

# Conclusions

For the first time to our knowledge, the PS corona structure was investigated with cryo-TEM over more than 2 400 NPs, allowing a statistical analysis. Our results show first that the corona is not limited to supported lipid bilayers, as sometimes assumed in the literature, but comprises a wide variety of intermediate structures. We thus provide a new definition of the PS corona: the pulmonary surfactant corona is the pattern constituted by the different hybrid structures formed by Nps and phospholipids and by their relative amounts. Second, we demonstrate that membrane and particle electrostatic charges control the interaction, and consequently the PS corona. These findings having significant implications to the pulmonary toxicity studies on nanoparticles, our results have now to be validated in animals. In addition, pre- or post- pulmonary surfactant treatment is shown to significantly affect the PS corona, which underline the need of protocols harmonization for sample preparation to avoid the physico-chemical or biological studies of unrelevant coronas.

Understanding the PS corona is crucial. First, the formation of this corona may interfere with the physiological function of the endogenous PS. Second, because the PS corona is formed at the initial biological barrier in the lungs, its biophysicochemical properties determine the subsequent biological identity and fate of the inhaled NPs. Thus, knowing the PS corona architecture will shed light on understanding molecular mechanism of nano-bio interactions in the respiratory system. Third, understanding the PS corona structure and formation is a way to intentionally formulate specific NPs / PS colloids effective as drug delivery carriers in lungs.[36]

**Conflicts of interest**
There are no conflicts to declare.

# Acknowledgements

ANR (Agence Nationale de la Recherche) and Solvay are gratefully acknowledged for their financial support of this work through ANR-12-CHEX-0011 (PULMONANO).

# Notes and references